\documentstyle[12pt]{article}
\begin{document}
\setcounter{section}{0}
\begin{center}
\Large\bf Some decay modes of the $1^{-+}$ hybrid meson 
in QCD sum rules revisited
\end{center}
\vspace{0.5cm}
\begin{center}
{ Shi-Lin Zhu}\\\vspace{3mm}
{Institute of Theoretical Physics \\
Academia Sinica, P.O.Box 2735\\
Beijing 100080, China\\
FAX: 086-10-62562587\\
TEL: 086-10-62569358\\
E-MAIL: zhusl@itp.ac.cn}
\end{center}
\vspace{1.0cm}
\begin{abstract}
The pionic coupling constants in the decays of the $1^{-+}$ hybrid 
meson are calculated. The double Borel transformation is invoked and 
continuum contribution is subtracted. 
The decay widths of the processes $1^{-+}\to \rho\pi , f_1 \pi, 
\pi \gamma$ are around $40, 100, 0.3$ MeV respectively.
Comparison is made with previous calculations using three point 
correlation functions.
\end{abstract}



\pagenumbering{arabic}

\vspace{1cm}

There appears increasing experimental evidence for a $J^{PC}=1^{-+}$ hybrid 
meson. E852 \cite{e852} and Crystal Barrel \cite{cb} collaboration reported 
a resonance with mass and width 
$1370\pm 16^{+50}_{-30}$MeV, $385\pm 40^{+65}_{-105}$MeV and 
$1400\pm 20 \pm 20$MeV, $310\pm 50^{+50}_{-30}$MeV both in $\eta \pi$ channel 
respectively. Beladidze et al. in VES experiment at IHEP reported a 
broad signal in the $\eta\pi^-$ state \cite{ves}. 
Very recently E852 collaboration observed a $J^{PC}=1^{-+}$ exotic state
with a mass of $1593\pm 8^{+29}_{-47}$ MeV and a width of $168\pm 20^{+150}_{-12}$ MeV 
\cite{prl-98} in the $\rho\pi$ channel in the reaction $\pi^- p\to \pi^+\pi^-\pi^- p$
at $18$ GeV.

We have calculated the binding energy and decay modes 
of heavy hybrid mesons with a heavy quark in the framework of heavy quark 
effective theory using the light cone QCD sum rule technique \cite{zhu-hybrid}. 
In this work we extend the same formalism to calculate the decay widths of 
the processes $1^{-+}\to \rho\pi , f_1 \pi, \pi \gamma$. 

Denote the isovector $J^{PC}=1^{-+}$ hybrid meson by ${\tilde \rho}$.
The interpolating current for ${\tilde \rho}$ reads
\begin{equation}
\label{cu1}
J_\mu (x) ={\bar u}(x)g_s \gamma^\nu G^a_{\mu\nu}(x)
{\lambda^a\over 2} d (x)\;,
\end{equation}
The overlapping amplitude $f_{\tilde \rho}$ is defined as
\begin{equation}\label{overlap-1}
\langle 0 | J_\mu (0) | {\tilde \rho}\rangle =\sqrt{2}f_{\tilde \rho}
m^3_{\tilde \rho}  \epsilon_\mu \; ,
\end{equation}
where $\epsilon_\mu$ is the ${\tilde \rho}$ polarization vectors. 

The decay amplitude for the p-wave decay process ${\tilde \rho} \to \rho \pi $ is
\begin{equation} \label{1-1}
 M({\tilde \rho}\to \rho \pi  )= i\epsilon_{\mu\alpha\sigma\beta}
\epsilon^\mu e^\alpha q^\sigma p^\beta g_1  \;,
\end{equation}
where $e_{\mu}$ is the polarization vector of the rho meson.

For the decay ${\tilde \rho \to f_1(1285) \pi}$, there exist two independent 
coupling constants, corresponding to S-wave and D-wave decays. Since the 
D-wave decay width is much smaller than S-wave width, we shall consider 
only the sum rules for the S-wave decay coupling constant.
The decay amplitude is:
\begin{equation} \label{1-5}
 M({\tilde \rho}\to f_1 \pi  )= (\eta\cdot\epsilon)g_2 +\cdots \;,
\end{equation}
where $\eta_{\mu}$ is the polarization vector of $f_1$ meson.

We consider the correlators 
\begin{equation}\label{cor-1}
i \int d^4x\;e^{ip\cdot x}\langle\pi (q)|T\left(J^\alpha_\rho (x)
J^{\dagger}_\mu(0)\right)|0\rangle 
=i\epsilon_{\mu\alpha\sigma\beta} q^\sigma p^\beta G_1 (p^2, p'^2) \; ,
\end{equation}
\begin{equation}\label{cor-2}
i \int d^4x\;e^{ip\cdot x}\langle\pi (q)|T\left(J^\alpha_{f_1}(x)
 J^{\dagger\mu}(0)\right)|0\rangle 
=  g^{\mu\alpha} G_2 (p^2, p'^2) +\cdots \;, 
\end{equation}
where $p'=p-q$, 
$J^\alpha_\rho (x)={1\over \sqrt{2}}[{\bar u}\gamma^\alpha u (x)-
{\bar d}\gamma^\alpha d (x)]$, 
$J^\alpha_{f_1} (x)={1\over \sqrt{2}}[{\bar u}\gamma^\alpha \gamma_5 u (x)+
{\bar d}\gamma^\alpha \gamma_5 d (x)]$,
$<0|J^\alpha_\rho |\rho > =f_\rho e^\alpha$, and 
$<0|J^\alpha_{f_1} |f_1 > =f_{f_1} \eta^\alpha$. 

Since the steps to derive the sum rule for the coupling constant
$g_{1,2}$ are very similar to those in \cite{zhu-hybrid}, we omit the details
and present the final results directly.
\begin{equation}\label{g1}
\sqrt{2}f_{\tilde \rho}m^3_{\tilde \rho}f_\rho m_\rho g_1
e^{-( { m_\rho^2 \over M_1^2}+{ m^2_{\tilde \rho} \over M_2^2} )}= 
-\sqrt{2}f_\pi \{ [\Phi_\bot (u_0)-{\tilde \Phi}_\bot (u_0)
+{\tilde\Phi}_{\|}(u_0)] M^2 + {1\over 36} <0|g_s^2 G^2|0> \phi_\pi (u_0) 
\} \;,
\end{equation}
\begin{equation}\label{g2}
\sqrt{2}f_{\tilde \rho}m^3_{\tilde \rho}f_{f_1} m_{f_1}  g_2
e^{-( { m_{f_1}^2 \over M_1^2}+{ m^2_{\tilde \rho} \over M_2^2} )}= 
{f_\pi\over \sqrt{2}}\{ [\Phi'_\bot (u_0)-2{\tilde \Phi}'_\bot (u_0)]
M^4 + {1\over 36} <0|g_s^2 G^2|0> \phi'_\pi (u_0) M^2 
\} \;,
\end{equation}
where $u_0={M^2_1 \over M^2_1 + M^2_2}$, 
$M^2 \equiv {M^2_1M^2_2\over M^2_1+M^2_2 }$, $M_1^2$, $M^2_2$ 
are the Borel parameters. The definitions of pion wave functions can be found
in \cite{zhu-hybrid,bely-95} and  $\Phi'_\bot (u) =
{d \Phi_\bot (u) \over du}$ etc. The sum rule is asymmetric
with the Borel parameter $M^2_1$ and $M^2_2$ since the hybrid meson is 
heavier than the rho or $f_1(1285)$ meson. For simplicity, we have 
given the expressions after integration of the double spectral density 
in the interval $(0, \infty)$ for the right hand side of (\ref{g1}) and
(\ref{g2}). The subtraction of the continuum contribution is discussed 
in \cite{zhu-hybrid}, which is crucial for the numerical analysis.

The values of the input parameters are $f_\pi=0.132$ GeV,
$m_{\tilde \rho}=1.6$ GeV, $f_{\tilde \rho}=0.026$ GeV \cite{book}, 
$m_\rho =0.77$ GeV, $f_\rho =0.22$ GeV, 
$m_{f_1}=1.285$ GeV, $f_{f_1}=0.24$ GeV. 
We have used the mass sum rules of $f_1 (1285)$ \cite{reinders} to 
obtain $f_{f_1}$. Moreover we use $\delta =0.18$ GeV$^2$ instead of
$\delta =0.2$ GeV$^2$ as in \cite{zhu-hybrid,bely-95}.

Let $M^2_1 =2\beta m_{\rho, f_1}^2$, $M^2_2=2\beta m^2_{\tilde \rho}$,
where $\beta$ is the dimensionless scale parameter.
Then we have $u_0 ={m_{\rho, f_1}^2\over m_{\rho, f_1}^2
+m^2_{\tilde \rho}}$, $M^2={2m_{\rho, f_1}^2 m^2_{\tilde \rho}\over 
m^2_{\tilde \rho}+ m_{\rho, f_1}^2}\beta$. 

The sum rules (\ref{g1}) and (\ref{g2}) is stable with 
reasonable variation of the Borel parameter 
$M^2$ and the continuum threshold $s_0$. 
In order to avoid the possible contamination from the radial excited
states $\rho (1450)$ and $f_1(1420)$ we choose the continuum 
$s_0 = (2.2\pm 0.2) $GeV$^2$. Numerically we have 
\begin{equation}
 g_1  = (2.6\pm 1.2)\mbox{GeV}^{-1}\;,
\end{equation} 
\begin{equation}
g_2  = (5\pm 2)\mbox{GeV}\;.
\end{equation}
The central value corresponds to $\beta =1.2$ and $s_0=2.2$ GeV$^2$.
The errors refers to the variations with $M^2$, uncertainty of $s_0$,
uncertainty of the pion wave functions, and the inherent uncertainty of 
the light cone qCD sum rule approach. Especially the sum rule for $g_2$
involves the first derivative of pion wave functions so it is less reliable
than that for $g_1$.

The coupling constant $|g_1|$ was first calculated to be around 
$2\sim 7$ GeV$^{-1}$ with $m_{\tilde \rho}=1.3$ GeV 
using three-point correlation functions at the symmetric point
$p^2 =q^2 =p'^2 $ in \cite{deviron-decay}. 
Later the hybrid mass and vertex sum rules were reanalysed 
leading to $g_1 =9 \sim 10 $ GeV$^{-1}$ \cite{zphys} and 
$7.7$ GeV$^{-1}$ \cite{book} with $m_{\tilde \rho}=1.5$ GeV.
The sum rules calculated at the symmetric point receive large 
contamination from the higher resonances and the continuum contribution
since only single Borel transformation can be invoked,
which renders its prediction less relaible. In order to illustrate
this point more clearly we let $s_0 \to \infty$, i.e., 
with the continuum contribution unsubtracted.
In this case we arrive at $g_1 =(5.2\pm 2.0)$ GeV$^{-1}$, 
which is numerically close to the value $g_1 =7.7$ GeV$^{-1}$ 
in \cite{book}. In other words, the continuum contributes as large 
as the ground state so its subtraction is cruicial for a reliable 
extraction of the coupling constant.

The formulas of the decay widths are
\begin{equation}
\Gamma ({\tilde \rho}^-\to \rho^-\pi^0 +\rho^0\pi^- )= {g_1^2\over 12\pi}
|{\vec q}_\pi |^3 \;,
\end{equation}
\begin{equation}
\Gamma ({\tilde \rho}^-\to f_1\pi^- )= {g_2^2\over 24\pi}
{|{\vec q}_\pi |\over m^2_{\tilde \rho} } 
(3+ {|{\vec q}_\pi |^2\over m^2_{f_1} }) \;,
\end{equation}
where $|{\vec q}_\pi |$ is the pion decay momentum.
Numerically,
\begin{equation}
\Gamma ({\tilde \rho}\to \rho\pi )= (40\pm 20)\mbox{MeV}\;,
\end{equation}
\begin{equation}
\Gamma ({\tilde \rho}\to f_1\pi )= (100\pm 50)\mbox{MeV} \;.
\end{equation}
We may further assume the vector dominance to relate the coupling 
constant for the process ${\tilde \rho}\to \gamma\pi$ to 
$g_1$ as in \cite{zphys},
$g_{{\tilde \rho} \gamma\pi } ={e\over 2\gamma_\rho} g_1\sim 0.15$ GeV$^{-1}$,
where $\gamma_\rho =2.56$. In this way we can estimate 
$\Gamma ({\tilde \rho}\to \gamma\pi ) \approx g^2_{{\tilde \rho} \gamma\pi }
{m^3_{\tilde \rho}\over 96\pi} \sim 300$ keV. The width of 
$\rho\pi$ decay channel from the present calculation is much 
smaller than those from the vertex sum rules, which is 
$600$ MeV \cite{zphys} and $250$ MeV \cite{book}.

One might take a step further and try to extend the same formalism to the
decay process $1^{-+}\to b_1 (1235) \pi$. However, the $b_1 (1235)$ mass 
sum rule is not stable \cite{reinders}. We do not consider this mode in 
this work. In short summary we have updated the QCD sum rule predictions 
for the pionic coupling constants in the light exotic meson decays and 
estimated the widths of some decay modes.

\vspace{0.8cm} {\it Acknowledgments:\/} This project was supported by
the Natural Science Foundation of China.

\end{document}